\documentclass{article}
\usepackage{amsmath,amssymb}

\begin{document}
\parskip=3pt
\baselineskip=22pt {\raggedleft{\bf XJUCTP-05002\\}}
\bigskip
\smallskip
\centerline{\Large\bf COUPLING CONSTANTS in ASYMPTOTIC EXPANSIONS
\vspace{2ex} \footnote{\sf 1 Supported by NSFC no.
10265003.}}\vspace{2ex} \vspace{4ex}

\centerline{\large\sf Chao-Zheng Zha\footnote{\sf Tel.:
86-10-60306582; e-mail address: czz@xju.edu.cn}} \vspace{3ex}
\centerline{\sf CCAST(World Laboratory)} \centerline{\sf P.O. Box
8730, Beijing 100080, China} \vspace{2ex} \centerline{\sf and}
\vspace{2ex} \centerline{\sf Department of Physics and Center for
Theoretical Physics} \vspace{2ex} \centerline{\sf Xinjiang
University, Wulumuqi Xinjiang 830046} \vspace{2ex} \centerline{\sf
People's Republic of China\footnote{\sf Mailing address.}}
\vspace{4ex}

\begin{center}
\begin{minipage}{5in}
\centerline{\large\bf Abstract} \vspace{3ex} {\sf ~~~~
Perturbation theory is a powerful tool in manipulating dynamical
system. However, it is reliable only for infinitesimal
perturbations except the first few approximations. We propose to
dispose this problem by means of perturbation group, and find that
the coupling constant approaches to zero in the limit of high
order perturbations. The Land\`{e} factor and the ground state
energy of one dimensional $\varphi^4$ theory are also concerned.}
\vspace{4ex}



\end {minipage}
\end{center}
\newpage
\section{Introduction}Perturbation theory is widely used in solving dynamical systems,
since generally we do not know how to deal with non-linear systems
directly \cite{kato,reed}. Although this method together with the
manipulation of the radiative corrections by renormalization
scheme has made prominent success of quantum electrodynamics in
the theoretical predictions for Lamb shift and anomalous magnetic
moment. However, early in 1952, Dyson \cite{dyson} argued that all
the asymptotic series used in quantum electrodynamics after
renormalization in mass and charge are divergent, and that the
origin in the complex plane of the coupling constant is a pole.
And even earlier, Titchmarsh \cite{tit} asserted that the
perturbation series are asymptotic series. In 1969, Bender and Wu
\cite{wu} discussed the $\varphi^4$-model in one dimension in
their pioneer works. They found that the ground state energy
$E_0(\lambda)$ is divergent, and there are an infinite sequence of
poles of the resolvent when the phase of the coupling constant
goes near to $\pm\dfrac{3}{2}\pi$, in addition to the cut from the
origin to $-\infty$ along the real axis pointed out by Jaffe
\cite{jaf}. In 1976, Lipatov \cite{lip} investigated the
renormalizable polynomial interaction scalar model. After
effecting the Watson-Sommerfeld transformation, he proved the
existence of the ultraviolet fixed point of the theory. Brezin et.
al. \cite{brezin}  re-derived the results obtained by Bender and
Wu \cite{wuba} in the model with internal $O(n)$ symmetry. In
2002, Kazakov and Popov \cite{kaza} showed that the asymptotic
series of the Gell-Mann-Low $\beta$-function cannot be recovered
by its first coefficients of the perturbation series and their
asymptotic values without invoking additional information. At the
same time, the mathematicians also paid great attention to
perturbation series. In 1967, Kato \cite{kato} set up a theorem
for the analytic behavior of of the eigenvalues of teh linear
operators in analytic family. Recently, Reed and Simon \cite{reed}
presented an systematic and extensive review for the theory of
perturbation. There the plague of the asymptotic series is
illustrated through concrete examples.

Now we propose to dispose this problem by means of perturbation
group method, explore the applications of perturbation group in
quantum field theory, and show that the coupling constant in the
perturbation series should be varied with respect to the order of
approximation as Dyson once expected. Finally,  in order to show
the availability of the perturbation group scheme, we take a
glance over the applications of perturbation group in anomalous
magnetic moment of electron, Gell-Mann-Low $\beta$-function and
the ground state energy of one dimensional $\varphi^4$-model.

\section{Perturbation Group}
It is well-known that the Rayleigh-Schr\"{o}dinger  series is only
the asymptotic series of the solution for the Schrodinger
equation. The series $\sum\limits_{n=0}^{N}a_n\lambda^n$ is
referred as an asymptotic to $f(\lambda)$ means that
\begin{equation}
\lim\limits_{\lambda\downarrow
0}\Big(f(\lambda)-\sum\limits_{n=0}^{N}a_n\lambda^n\Big)\Big/\lambda^N=0
\end{equation}
It is shown that if the asymptotic series is not convergent, then
it is reliable only when $\lambda$ is in the neighbor of zero
\cite{kato,wu,reed}, otherwise it gives no reliable information
for  the values of the function $f(\lambda)$ at some finite value
of $\lambda$ as one might expect. Therefore the perturbation
series is reliable only  for infinitesimal perturbations rather
than any finite perturbation except the first few approximations.
In view of improving the exactness and reliableness of the
perturbation method , it is preferable to separate the
perturbation into many steps and limit the perturbations to be
infinitesimal in each step.

Let $H_0$ and $K$ be two elements in a Banach space ${\cal B}$,
and $\beta\in \mathbb{Z}$, then
$$
H=H_0+\beta K\in {\cal B}.
$$

Let $T(\beta,\beta_0)$  be a translation in ${\cal B}$,
\begin{eqnarray}
T(\beta, \beta_0):~~~~H_{\beta_0}=H_0+\beta_0 K\longmapsto
H_{\beta}=H_{0}+\beta K.
\end{eqnarray}

Let the representation $U(\beta,\beta_0)$ of the translation
$T(\beta, \beta_0)$ be a transformation on a Hilbert space ${\cal
H}$ defined by $U(\beta_1, \beta_0)\in{\cal L}({\cal H})$,
\begin{eqnarray}
U(\beta_1, \beta_0)\varphi_{\beta_0}=\varphi_{\beta_1}
\end{eqnarray}
where $\varphi_{\beta_0}$ and $\varphi_{\beta_1}$ are the
eigenvectors of $H_{\beta_0}$ and $H_{\beta_1}$ respectively. And
it is assumed that when $\beta$ varies continuously from $\beta_0$
to $\beta_1$, the eigenvector of  $H(\beta, \beta_0))$ varies
continuously from $\varphi(\beta_0)$ to $\varphi(\beta_1)$. Since
\begin{eqnarray}
U(\beta_2, \beta_1)U(\beta_1, \beta_0)\varphi_{\beta_0}=U(\beta_2,
\beta_1)\varphi_{\beta_1}=\varphi_{\beta_2},
\end{eqnarray}
and
\begin{eqnarray}
U(\beta_2, \beta_0)\varphi_{\beta_0}=\varphi_{\beta_2},
\end{eqnarray}
Thus
\begin{eqnarray}
U(\beta_2, \beta_1)U(\beta_1, \beta_0)=U(\beta_2, \beta_0).
\end{eqnarray}
Besides
\begin{eqnarray}
U(\beta, \beta)=1,
\end{eqnarray}
and
\begin{eqnarray}
U(\beta_1, \beta_0)^{-1}=U(\beta_0, \beta_1).
\end{eqnarray}
Therefore the transformations $U(\beta_1, \beta_0)$ form a group.

\section{Generator for Perturbation Group}
Let us first setup the perturbation equation for the perturbation
group. The derivative of $U(\beta, \beta_0)$ with respect to
$\beta$ is given as follows,
\begin{eqnarray}
\dfrac{\partial U(\beta,
\beta_0)}{\partial\beta}=&&\lim\limits_{\Delta\beta\rightarrow
0}\dfrac{U(\beta+\Delta\beta, \beta_0)-U(\beta,
\beta_0)}{\Delta\beta}\nonumber\\
=&&\lim\limits_{\Delta\beta\rightarrow
0}\dfrac{U(\beta+\Delta\beta, \beta)-U(\beta,
\beta)}{\Delta\beta}U(\beta, \beta_0)\nonumber\\
=&&\dfrac{\partial U(\beta', \beta)}{\partial
\beta'}\Bigg|_{\beta'=\beta}U(\beta, \beta_{0}).
\end{eqnarray}
where we denote
\begin{eqnarray}
G(\beta)\equiv\dfrac{\partial U(\beta', \beta)}{\partial
\beta'}\Bigg|_{\beta'=\beta}
\end{eqnarray}
as the generator of the perturbation group. thus we have the
perturbation equation for the perturbation group as follows,
\begin{eqnarray}
\dfrac{\partial U(\beta, \beta_0)}{\partial\beta}=G(\beta)U(\beta,
\beta_{0}).
\end{eqnarray}

The infinitesimal transformation can be written as
\begin{eqnarray}
U(\beta+\Delta\beta, \beta)=\exp\{ G(\beta)\Delta\beta\},~~~~ {\rm
for}~~~~\Delta\beta\rightarrow 0
\end{eqnarray}
Therefore we can proceed along with the infinitesimal
perturbations step by step, and finally obtain the perturbation
with  finite coupling constant. Let $C$ be a section of line
connecting two points $\beta_0$ and $\beta_1$ in the analytic
region of the complex plane of $\beta$, while the line segment is
divided into $n$ sections by $n-1$ points
$\beta(s_k),s_k\in\mathbb{R},(k=1,2,3,....,n-1)$ on the line,
\begin{eqnarray}
\beta(s_k)=\beta(s_0)+k\Delta\beta,~~(k\in\mathbb{I})~~~~\beta(s_n)=\beta(s_f).
\end{eqnarray}
Then
\begin{eqnarray}
&&U(\beta_{f},\beta_0)=\lim\limits_{n\rightarrow\infty}U(\beta_0+n\Delta\beta,
\beta_0+(n-1)\Delta\beta) ....U(\beta_{0}+k\Delta\beta,
\beta_{0}+(k-1)\beta))...U(\beta_{0}+\Delta\beta,
\beta_0)\nonumber\\
&&=\lim\limits_{n\rightarrow\infty}\mathbb{P}\exp\{G(\beta_0+(n-1)\Delta\beta)\Delta\beta
\}...\exp\{G(\beta_{0})
\Delta\beta \}\nonumber\\
&&=\mathbb{P}\exp\{\lim\limits_{n\rightarrow\infty}\sum\limits_{k=0}^{n-1}G(\beta+k\Delta\beta)\Delta\beta)\}\nonumber\\
&&=\mathbb{P}\exp\Big\{\int_{\beta_0}^{\beta_f}G(\beta)d\beta\Big\}\nonumber\\
&&=1+\sum\limits_{n=0}^{\infty}\dfrac{1}{n!}\int\limits_{\beta_0}^{\beta_f}d\beta_{1}
...\int\limits_{\beta_0}^{\beta_f}d\beta_{n}\mathbb{P}
G(\beta_1)G(\beta_2)...G(\beta_n). \label{u}
\end{eqnarray}
where $\mathbb{P}$  denotes the perturbation ordered product by
the coupling constant $\beta\in\mathbb R$,
\begin{eqnarray}
\mathbb{P}G(\beta(s_2))G(\beta(s_1))=\Big\{\begin{array}{cc}G(\beta(s_2))G(\beta(s_1)),&\mbox{
for} s_2>s_1,\\
G(\beta(s_1))G(\beta(s_2)),&\mbox{for} s_1>s_2,
\end{array}
\end{eqnarray}
and use is made of the fact that
\begin{eqnarray}
&&\mathbb{P}\int\limits_{\beta_0}^{\beta_f}d\beta_{1}
...\int\limits_{\beta_0}^{\beta_{n-1}}d\beta_{n}
G(\beta_1)G(\beta_2)...
G(\beta_n)=\dfrac{1}{n!}\int\limits_{\beta_0}^{\beta_f}d\beta_{1}
...\int\limits_{\beta_0}^{\beta_f}d\beta_{n}\mathbb{P}
G(\beta_1)G(\beta_2)...G(\beta_n).
\end{eqnarray}

Note that in view of the of the perturbation ordered product,
there are no commutators between the generators in the exponential
in contrast to the Campbell-Baker-Hausdorff formula.

The transformation $U(\beta_f,\beta_0)$ in eq.(\ref{u}) can be
manipulated through iteration, i. e., insert the zeroth
approximation into the right hand side integrals and obtain the
first approximation, then iterate. In time dependent quantum
mechanics, the perturbation transformation $U$ and thus the
generators are also time dependent.

\section{Perturbation Group in Quantum Field Theory}
The evolution of the state vector $\varphi_I(\beta; t)$ in the
interaction picture is effected by the evolution operator
$U(\beta, \beta_0; t_1, t_0)$,
\begin{eqnarray}
i\dfrac{\partial U(\beta_0+\Delta\beta, \beta_0; t ,t_0)}{\partial
t}&&= \Delta\beta H_{I}(t)U(\beta_0+\Delta\beta, \beta_0; t ,t_0),
\end{eqnarray}
where $H_{I}(t)\equiv\int d^3x{\cal H}(t, \mathbf{x})$ is the
Hamiltonian 3-density, and
\begin{eqnarray}
U(\beta_0+\Delta\beta, \beta_0; t ,t_0)=\mathbb{T}\exp\{i\int dt
\Delta\beta H_I(t)\}
\end{eqnarray}
It can be easily seen that the perturbation transformations form a
group. While the generator for the infinitesimal perturbation
transformation is
\begin{eqnarray}
G(\beta)=i\int dtH_I(t),
\end{eqnarray}

Then the perturbation group equation is
\begin{eqnarray}
\dfrac{\partial U(\beta_1, \beta_0; t,
t_0)}{\partial\beta}=\Big\{\int dtH_I(t)\Big\}U(\beta_1, \beta_0;
t, t_0)
\end{eqnarray}

Therefore the  perturbation transformation $U(\beta, 0; t ,t_0)$
can be obtained  as follows,
\begin{eqnarray}
U(\beta, 0; t,
t_0)&&=\mathbb{PT}\exp\Bigg\{-i\int\limits_{0}^{\beta}
d\beta'\int_{t_0}^{t}dtH_{I}(t')\Bigg\}\nonumber\\
&&=1+\sum\limits_{n}\dfrac{1}{n!}\dfrac{(-i)^n}{n!}\beta^n\int\limits_{t_0}^tdt'_1...\int\limits_{t_0}^tdt'_n
\mathbb{T}H_I(t'_1)...H_I(t'_n).
\end{eqnarray}

In view of the reducing factor $\dfrac{1}{n!}$ before the n-th
order term in the series, it seems that the contributions of the
higher corrections are overestimated in the conventional
series.\cite{kaza}. This reducing factor $\dfrac{1}{n!}$ can also
be absorbed into the coupling constant, and define the coupling
constant as a function $g(n)$ of order of approximation $n$,
\begin{eqnarray}
g(n)\equiv\dfrac{g}{(n!)^{1/n}}\approx\dfrac{e}{n}g,
\end{eqnarray}
for large $n$, where  $e$ is the constant of the natural
logarithm. Therefore
\begin{eqnarray}
\lim\limits_{n\rightarrow\infty}g(n)=0,
\end{eqnarray}
as Dyson once expected \cite{dyson}. Then the series recovers its
conventional form.

\section{Quantum electrodynamics, quantum chronodynamics and $\varphi^4$ theory}

It is well-known that quantum electrodynamics achieved great
success in theoretical prediction for anomalous magnetic moment of
electron\cite{hughes}. According to Furry's theorem, it seems to
be more reasonable to choose the coupling constant $\alpha\equiv
e^2$ as the parameter in loop expansions.

The Land\`{e} factor $g$ for anomalous magnetic moment of pure
quantum electrodynamics in perturbation group scheme \cite{hughes}
is
\begin{eqnarray}
g=2\Big[1+C_1\Big(\dfrac{\alpha}{\pi}\big)+C_2\dfrac{1}{2!}\Big(\dfrac{\alpha}{\pi}\big)^2
+C_3\dfrac{1}{3!}\Big(\dfrac{\alpha}{\pi}\big)^3
+C_4\dfrac{1}{4!}\Big(\dfrac{\alpha}{\pi}\big)^4+...\big],
\label{g}
\end{eqnarray}
where the coefficients $C_i$'s are obtained in renormalization.

Taking the contributions from QED into account only, the
$g$-factor obtained in perturbation group approach is
\begin{eqnarray}
g^{\rm PG}_{\rm QED}=2(1+0.001 160 526 040 2..),
\end{eqnarray}
while recent experimental data is \cite{hughes}
\begin{eqnarray}
g_{\exp}=2(1+0.001 159 652 188 4...).
\end{eqnarray}

It is readily seen that $g^{\rm PG}_{\rm QED}$ is a little bit
closer to $g_{\exp}$ than $g^{\rm ord}_{\rm
QED}=0.00116584705(18)$ obtained in the ordinary approach
\cite{hughes} up to the fourth order of $\alpha$. Besides, in
ordinary manipulations one might still worry about the higher
terms may spoil the convergence of the asymptotic series
\cite{dyson}, while the situation in the perturbation group scheme
is substantially improved.

In quantum chronodynamics, the Gell-Mann-Low $\beta$-function in
the perturbation scheme can be defined by the following series
\cite{muta},
\begin{eqnarray}
\beta(g)=-\beta_0g^3-\beta_1\dfrac{1}{2!}g^5-\beta_2\dfrac{1}{3!}g^7+O(g^9),
\end{eqnarray}
Thus it is evident that the behavior of the ordinary
$\beta$-function is substantially modified by the factors $1/n!$
in the perturbation group scheme. And this is also true for the
ground state energy $E_0(\lambda)$ of one dimensional
$\varphi^4$-model \cite{wu}.

Actually, the Hamiltonian of the one-dimensional anharmonic
oscillator is
\begin{eqnarray}
H=\dfrac{1}{2}\dot{\varphi}^2+\dfrac{1}{2}m^2\varphi^2+\lambda\varphi^4.
\end{eqnarray}
Then the ground state energy of the oscillator as observed by
Bender and Wu \cite{wuba} is given by The ground state energy of
the anharmonic oscillator is
\begin{eqnarray}
E_0(\lambda)=\dfrac{1}{2}m+\sum\limits_{n=1}^{\infty}mA_n(\lambda/m^3)^n.
\end{eqnarray}

The detail asymptotic growth of $A_n$ is
\begin{eqnarray}
A_n\sim(-1)^{n+1}(6/\pi^3)^{1/2}\Gamma(n+\frac{1}{2})3^n.
\end{eqnarray}

In using the Stirling's formula, the perturbation group scheme
gives out the asymptotic values of coefficients $A_n$  as follows,
\begin{eqnarray}
A_n\sim&&(-1)^{n+1}\Big(\dfrac{6}{\pi^3}\Big)^{1/2}\dfrac{\Gamma\big(n+\dfrac{1}{2}\big)3^n}{n!}\\
\sim&&(-1)^{n+1}\Big(\dfrac{6}{\pi^3}\Big)^{1/2}\dfrac{2e3^n}{(n+1)^{3/2}}
\end{eqnarray}
where $e$ is the constant for natural logarithm. It can be readily
seen that when $m=1$ and $0<\lambda<\dfrac{1}{3}$, and $E_0$ will
approach to a finite value when $n\rightarrow\infty$. When
$\lambda=0.2$, $E_0\approx1.131568$.

\section{Conclusion}
We propose the perturbation group and perturbation  equation.
Using the Rayleigh-Schrodinger series, we derive the generator for
perturbation transformations, give out the transformation for
finite perturbations, and find that the coupling constant in
quantum field theory is varied and approaches to zero as the order
of approximation goes to infinity in ordinary perturbation series
as Dyson once expected. It is shown that the prediction for
Land\`{e} factor in perturbation scheme matches the experiments
better than that obtained in ordinary manipulations, and the
behaviors of the asymptotic series in quantum field theories are
substantially modified.

It is worthy to mention that perturbation group is different from
renormaliztion group. In renormaliztion group, the coupling
constant is related to the energy scale of the renormalization
point. But in perturbation group, the coupling constant is
referred as an independent variable. Actually the final limits of
the coupling constant variables in  perturbation group
transformations are just the running coupling constants in the
perturbation series in physics. If one manipulates only by means
of renormaliztion group, then the asymptotic series in quantum
field theory are in general divergent even after renormalization
whenever the perturbations are non-infinitesimal. While the
behaviors of the asymptotic series are substantially modified in
perturbation group scheme.

\vspace{2ex}
\bigskip
\smallskip
{\raggedright{\large \bf Acknowledgement}}\bigskip ~~~~The author
would like express his hearty thanks to the Center of Cross
Disciplines in Science sponsored by Professor L. Yu for the
hospitality during the completion of part of this work. The author
would also like to thank Professors H.-Y. Guo, K. Wu and W.-Z.
Zhao for helpful discussions.

\end{document}